\documentclass[10pt,conference]{IEEEtran}

\ifCLASSINFOpdf
\else
\fi

\IEEEoverridecommandlockouts
\ifCLASSOPTIONcompsoc
  \usepackage[caption=false,font=normalsize,labelfont=sf,textfont=sf]{subfig}
\else
  \usepackage[caption=false,font=footnotesize]{subfig}
\fi
\usepackage{xcolor}
\usepackage{lipsum}
\usepackage[T1]{fontenc}
\usepackage{amsmath,amsfonts}
\usepackage{tikz-cd}
\usepackage{amssymb}
\usepackage{tikz}
\usepackage{mathrsfs}
\usetikzlibrary{positioning, shapes, arrows.meta}
\usepackage{algorithm}
\usepackage{algpseudocode}
\usepackage{array}
\usepackage{makecell}
\usepackage{amsthm}
\usepackage{adjustbox}
\usepackage{stackengine}
\usepackage{tabularx}
\usepackage{amssymb}
\usepackage{array}

\newcolumntype{P}[1]{>{\centering\arraybackslash}p{#1}}
\newcolumntype{M}[1]{>{\centering\arraybackslash}m{#1}}

\usepackage[colorlinks,
linkcolor=red,
anchorcolor=blue,
citecolor=green
]{hyperref}
\usepackage{booktabs}

\usepackage{amsmath} 
\usepackage{xcolor}
\usepackage{textcomp}
\usepackage{stfloats}
\usepackage{amsmath}
\usepackage{url}
\usepackage{verbatim}
\usepackage{graphicx,booktabs,multirow,bm}
\usepackage{subfig}
\usepackage{xcolor}
\usepackage{scalerel}
\usepackage{cite}
\usepackage{colortbl}

\begin{document}

\title{\huge Mutual Information-Empowered Task-Oriented Communication: Principles, Applications and Challenges}

\author{\IEEEauthorblockN{Hongru~Li, Songjie~Xie, Jiawei~Shao, Zixin~Wang, Hengtao~He\\ Shenghui~Song, Jun~Zhang and Khaled B.~Letaief \thanks{Hongru~Li, Songjie~Xie, Zixin~Wang, Hengtao~He, Shenghui~Song, Jun~Zhang and Khaled B.~Letaief are with the Department of Electronic and Computer Engineering, The Hong Kong University of Science and Technology (HKUST), Hong Kong, China. (e-mail: \{hlidm, sxieat\}@connect.ust.hk, \{eewangzx, eehthe, eeshsong, eejzhang, eekhaled\}@ust.hk). 
Jiawei~Shao is with the Institute of Artificial Intelligence (TeleAI), China Telecom, China (email: shaojw2@chinatelecom.cn) (The corresponding author is Hengtao He.)
}\\
}}

\maketitle

\pagestyle{plain}  
\thispagestyle{empty}

\begin{abstract}
Mutual information (MI)-based guidelines have recently proven to be effective for designing task-oriented communication systems, where the ultimate goal is to extract and transmit task-relevant information for downstream task. This paper provides a comprehensive overview of MI-empowered task-oriented communication, highlighting how MI-based methods can serve as a unifying design framework in various task-oriented communication scenarios. We begin with the roadmap of MI for designing task-oriented communication systems, and then introduce the roles and applications of MI to guide feature encoding, transmission optimization, and efficient training with two case studies. We further elaborate the limitations and challenges of MI-based methods. Finally, we identify several open issues in MI-based task-oriented communication to inspire future research.

\end{abstract}

\section{introduction}
\label{sec:introduction}
As Shannon and Weaver pointed out in their seminal work~\cite{shannon1948mathematical}, communication problems can be categorized into three levels: (1) technical problems, which focus on the accurate transmission of symbols; (2) semantic problems, which concern the interpretation of transmitted information; and (3) effectiveness problems, which evaluate whether the received information successfully achieves the intended goal. Traditional communication systems have largely focused on solving technical problems, ensuring high-fidelity transmission of raw data. However, in the era of big data and artificial intelligence (AI), the sheer volume of transmitted data makes the traditional design highly inefficient. This has led to an increasing demand for addressing semantic and effectiveness problems, where the focus shifts from transmitting raw data to delivering only task-relevant information. In this context, task-oriented and semantic-aware communication~\cite{vfe, xie2021deep}, has emerged as a key enabler for next-generation communications, optimizing transmission based on the specific needs of downstream tasks rather than ensuring bit-level accuracy.

A fundamental challenge in task-oriented communication is designing encoding and decoding mechanisms that prioritize the most relevant information for a given task~\cite{gunduz2022beyond}. Unlike traditional communication systems that focus on minimizing bit errors, task-oriented communication systems must ensure that only the essential information is transmitted, reducing redundancy and improving efficiency. However, achieving these goals requires a systematic approach for feature selection and transmission optimization, especially under dynamic and noisy wireless environments where signal distortions and distribution shifts can degrade system performance. Thus, a key research question is how to design a unified framework that optimizes communication efficiency and task-specific performance.

A promising approach to address this question is leveraging mutual information (MI) as a theoretical foundation for designing task-oriented communication systems. 
MI quantifies how much useful information is preserved and transmitted, making it a powerful metric for guiding encoding, feature selection, and transmission strategies. The pioneering work~\cite{xie2021deep} introduced MI to optimize data transmission rate, demonstrating that optimizing for semantic-level accuracy can significantly enhance system efficiency, but the learning performance and transmission optimization are not jointly considered. Building upon this, the authors of~\cite{vfe} first introduced a unified MI-based optimization framework, leveraging the information bottleneck (IB) principle, into task-oriented communication systems. This framework explicitly models the trade-off between communication efficiency and learning performance. As a result, it achieves a better rate-distortion trade-off compared to other separated optimization methods. The IB approach was later extended to multi-device cooperative task-oriented communication~\cite{shao2022task} with a distributed information bottleneck (DIB) principle, demonstrating that MI-based methodologies can generalize across diverse network environments and learning paradigms. To adapt to digital modulaiton and ensure robust transmission, the authors of~\cite{xie2022robust_IB} proposed a robust information bottleneck (RIB) framework, which integrates channel resilience directly into its MI-based optimization. This approach ensures that modulated symbols retain structured redundancy: sufficient information to recover task-critical features under channel distortions, but without extraneous data that would violate bandwidth or compatibility requirements. Furthermore, the authors in~\cite{cai2025end} proposed a MI-based optimization framework for precoding design in MIMO systems, which can jointly optimize the feature encoding, MIMO precoding and inference. These approaches show that MI-based frameworks provide an efficient foundation for encoding and decoding strategies that enhance both communication efficiency and learning performance.

Despite the promising theoretical advancements, there has yet to be a comprehensive review of MI-based methods for task-oriented communication. Furthermore, MI-based frameworks have been individually explored for feature selection~\cite{vfe}, transmission efficiency~\cite{cai2025end}, and robustness~\cite{xie2022robust_IB}, but no systematic survey integrates these contributions into a unified perspective. This paper aims to fill this gap by providing a comprehensive overview of MI-based methods in task-oriented communication, highlighting how these methods can be systematically connected through a common design pipeline—namely, probabilistic modeling, objective formulation, and learning algorithm design—and leveraged to address diverse system requirements such as adaptivity, efficiency, and robustness across different scenarios.

\begin{figure*}[t]
	\centering
        \includegraphics[width=\linewidth]{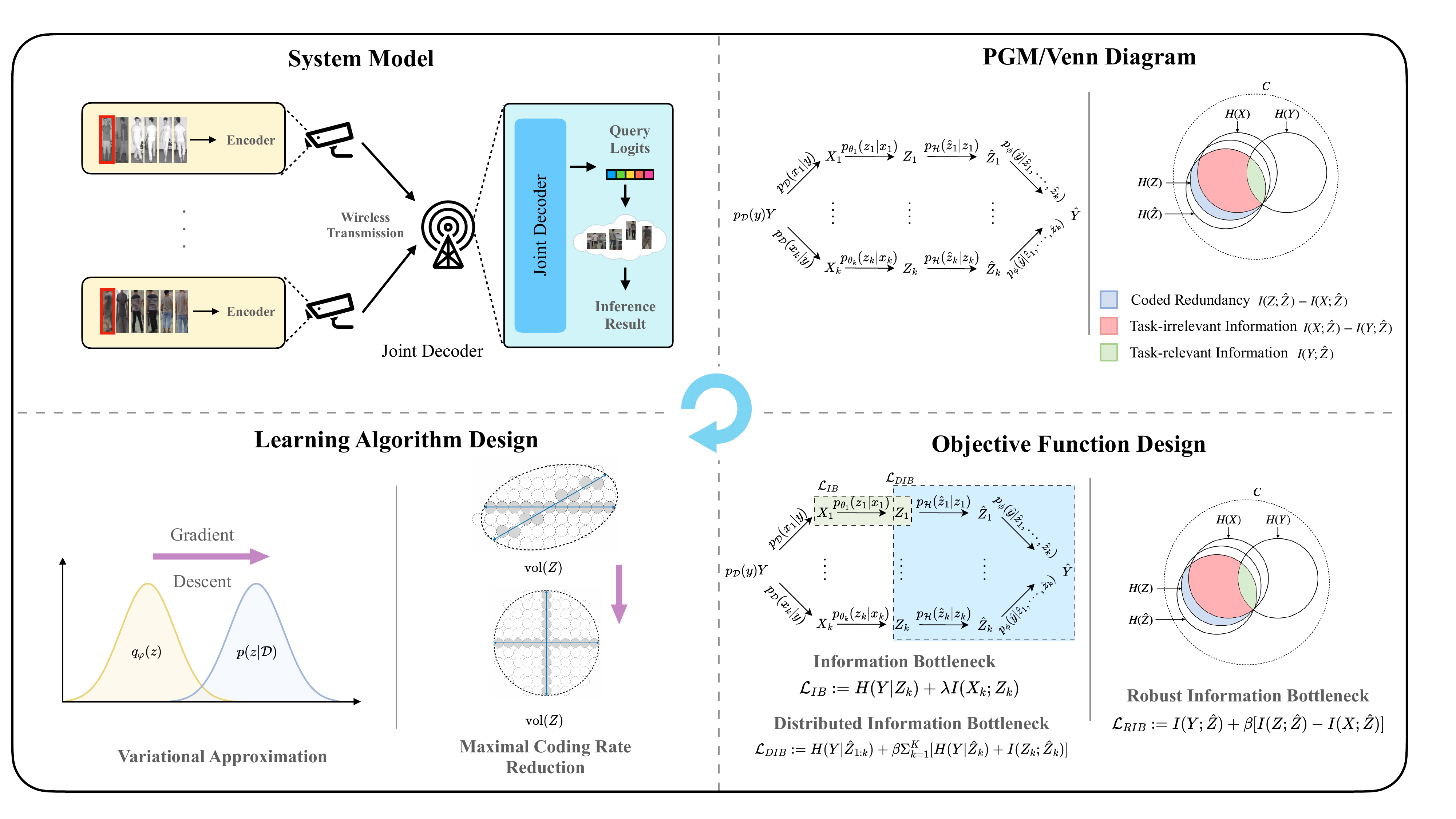}
        \caption{Three fundamental components of MI-based task-oriented communication system design.}
        \label{fig:component}
\end{figure*}

\section{Fundamental of Mutual Information for Task-Oriented Communication}
\subsection{Why Mutual Information?}

Traditional heuristic task-oriented communication designs often treat the communication and learning as separate processes, which are analyzed and optimized under different frameworks. The communication module typically prioritizes transmission efficiency, focusing on reconstructing intermediate features or logits at the receiver side. On the other hand, the learning module is independently designed to extract relevant features and make inference. This separated design leads to suboptimal performance, as learning objectives are not fully aligned with transmission strategies. MI provides a direct measure of information relevance, allowing for a joint optimization of both communication and learning objectives within a unified framework. Unlike traditional approaches that separately optimize communication efficiency (e.g., minimizing bit errors) and learning performance (e.g., improving inference accuracy), MI-based design explicitly models their interdependence. By optimizing MI-related objectives, task-relevant features are prioritized during encoding and transmission, ensuring that communication strategies are directly aligned with learning goals.

In information theory, MI is a fundamental concept that quantifies the mutual dependence between two random variables, making it a powerful tool for understanding dependencies in probabilistic models. By leveraging MI, \textit{\textbf{we can directly measure the mutual dependence between any two variables in a task-oriented communication system}}. MI provides a common metric to quantify both \textbf{task-relevance} and \textbf{transmission efficiency}, providing a unified framework for communication and learning optimization. For instance, the authors in~\cite{xie2022robust_IB} and~\cite{cai2025end} utilized MI to directly model the relationship between learning performance and key communication components such as modulation schemes and precoding matrices. This approach significantly improves system efficiency and robustness by ensuring that task-relevant information is preserved and optimized throughout the entire communication pipeline.

Furthermore, task-oriented communication lies at the intersection of communication and machine learning—two fields that have been extensively studied and have well-established theoretical foundations through the lens of MI. Given that MI serves as a common analytical tool in both domains, it provides a natural bridge for designing task-oriented communication systems. By incorporating MI into the system design, we can achieve a more holistic optimization that seamlessly integrates communication efficiency and learning performance, ultimately leading to more robust and efficient task-oriented communication systems.

\subsection{The Mutual Information-Based Design Framework}
As shown in Fig.~\ref{fig:component}, given a system model, we advocate for a structured workflow based on MI for designing task-oriented communication systems: 1) constructing a probabilistic graphical model (PGM) or Venn diagram from the system model, 2) formulating an MI-based objective function to optimize key system components, and 3) developing an efficient learning algorithm to estimate and optimize MI-based objectives. 

\textbf{1) PGM/Venn diagram modeling}: Since MI is defined in the probability space, constructing a PGM is essential for applying MI-based methods. The PGM represents the probabilistic dependencies between system variables, making it a powerful tool for modeling information flow in task-oriented communication systems. A key advantage of using PGM is that it allows for structured factorization of joint distributions, simplifying MI-based objective formulations. Within this framework, the $d$-separation principle provides a systematic way to analyze conditional independence relationships among variables. This is particularly useful when designing MI-based objectives, as it helps determine which system components contribute directly to MI calculations and which can be marginalized, and thus reducing computational complexity~\cite{li2024tackling}. In addition, Venn diagrams provide a visual representation of entropy and MI. In Fig.~\ref{fig:component}, each circle represents the entropy of a random variable, while the overlapping region represents the MI between variables, and indicates the shared information. For example, in feature encoding, the overlap between input features and task labels corresponds to task-relevant information, while non-overlapping regions indicate irrelevant or redundant information. By combining PGM for structured modeling and Venn diagrams for intuitive visualization, MI-based frameworks can better analyze and optimize information flow in task-oriented communication.

\textbf{2) Objective Functions Design}: With the established PGM and Venn diagram, MI-based objective functions guide the optimization of feature encoding, transmission, and inference. System parameters, such as encoder weights, precoding matrices, and modulation schemes~\cite{vfe, cai2025end, xie2022robust_IB} are modeled as the parameters of the transition probabilities in the MI thus we can directly optimize these parameters through MI-based objectives. 

\textbf{3) Learning Algorithm Design}: Direct computation of MI is often infeasible due to its high-dimensional distribution. To address this challenge, a dedicated learning algorithm is required to estimate MI-based objective funciton or find the suitable upper and lower bounds for efficient optimization. Various techniques, such as variational approximation~\cite{vfe}, neural MI estimation~\cite{li2024task}, and coding rate approximation~\cite{cai2025end} are widely used. These methods approximate or estimate MI-based objective function in an efficient way, enabling its integration into practical learning frameworks.


By integrating these three components, the principled and adaptable design framework of MI-based task-oriented communication system is established.
Next, we will elaborate several applications of MI-based methods in task-oriented communications 

\begin{figure*}[t]
	\centering
        \includegraphics[width=\linewidth]{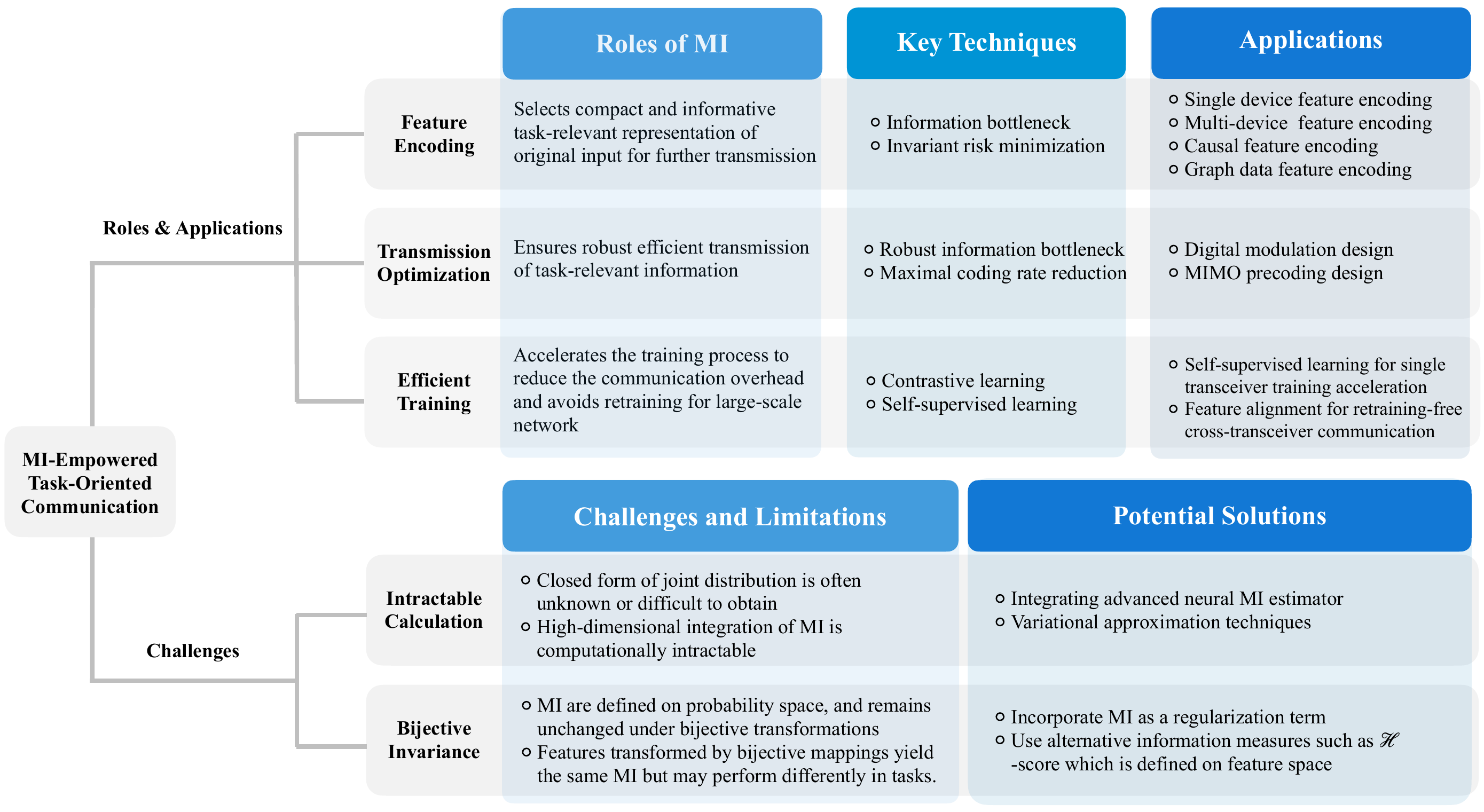}
        \caption{The roles, applications, and challenges of MI-based task-oriented communication.}
        \label{fig:overview}
\end{figure*}
\section{MI for Task-Oriented Communication: Roles, Applications, and Challenges}
This section elaborates the key roles, applications and challenges of MI in task-oriented communication, focusing on its applications in: feature encoding, which ensures that only essential task-relevant information is extracted and transmitted; transmission optimization, which adapts transmission strategies to preserve task-relevant features under different channel conditions; and efficient training, reducing communication overhead during learning while improving generalization. Additionally, we discuss the challenges associated with MI-based methods, including high-dimensional statistical estimation and the implications of its invariance to bijective transformations, which influence feature representation and inference performance.

By examining these aspects, we highlight both the strengths and limitations of MI-based approaches, offering insights into how they can be leveraged to design more efficient and adaptive task-oriented communication systems.

\subsection{Feature Encoding}
Feature encoding is a crucial component of task-oriented communication, as it determines which parts of the input data are retained and transmitted to maximize task performance while minimizing communication overhead. MI provides a principled approach to optimizing feature encoding by simultaneously quantifying the informativeness and compactness of the encoded features. Based on this idea, a series of representative schemes with IB principle have been proposed for task-oriented communication, including vanilla IB, distributed IB, invariant IB, and graph IB.

\subsubsection{Vanilla Information Bottleneck}
The IB principle formulates feature encoding as a Lagrange optimization problem that balances compression and relevance. PGM in this setting captures the dependencies between input data, encoded features, and task labels, ensuring that only task-relevant information is retained. The objective function follows the standard IB formulation: maximizing MI between encoded features and task labels while minimizing MI between encoded features and raw input to remove redundancy. To solve this optimization problem, a variational feature encoding (VFE) scheme is proposed in~\cite{vfe} to optimize the feature encoding process for task-oriented communication. To exhaust the compression potential of IB, a feature pruning algorithm is also proposed in~\cite{vfe} to reduce the Baud rate under dynamic channel conditions. As demonstrated in~\cite{vfe}, VFE significantly outperforms traditional non-MI-based methods and achieves a better rate-distortion trade-off, thereby highlighting the effectiveness of MI-based feature encoding in task-oriented communication.



\subsubsection{Distributed Information Bottleneck}
Building upon the vanilla IB framework, the authors in~\cite{shao2022task} adopt the DIB principle to enable multi-device cooperative task-oriented communication. In this scenario, PGM models the dependencies between observations from multiple devices, ensuring that each device transmits complementary, non-redundant information. The objective function introduces a distributed MI-based optimization, where spatial redundancy is minimized and task-relevant information is preserved. To optimize this objective, a deterministic distributed information bottleneck (DDIB) is proposed, which allows for flexible control of communication overhead among multiple devices. By leveraging MI as an optimization metric, DDIB can directly minimize the spatial redundancy and preserve complementary task-relevant information among different devices. The results in~\cite{shao2022task} demonstrate that DDIB outperforms traditional non-MI-based methods in terms of rate-distortion tradeoff on multi-device cooperative task-oriented communication systems, which further highlights the effectiveness of MI for feature encoding in distributed systems.




\subsubsection{Invariant Information Bottleneck}
In real-world scenarios, task-oriented communication systems must operate across diverse domains and adapt to unforeseen environmental shifts. A major challenge is ensuring that the encoded features remain robust to these variations while still preserving task-relevant information. To address this challenge, the authors in~\cite{li2024tackling} establish a PGM that captures the relationships among input data, ground truth, causal features, spurious features, and domain variables. Based on this structured representation, they propose invariant feature encoding (IFE) objective funciton based on invariant IB (IIB) that penalizes the MI between domain variables and ground truth, conditioned on the encoded features. This approach encourages the learning of causality-aware representations that remain invariant to domain shifts while filtering out task-relevant but domain-specific information. Additionally, an out-of-distribution (OOD) detection algorithm based on IB~\cite{li2024tackling} is introduced to identify and reject input data that deviates from the training distribution. This further enhances the robustness and reliability of the system under unseen conditions.

\subsubsection{Graph Information Bottleneck}
The irregular structure and large-scale nature pose significant challenges for feature encoding in graph data. To address this challenge, a graph IB (GIB) framework is proposed in~\cite{li2024task}. The PGM here captures the dependencies among subgraph structures, node attributes, and task labels, ensuring that both attribute-based and structural information are efficiently encoded. The objective function follows the IB formulation but applies MI constraints to both node-level features and graph topology. The optimization of IB-related objective always involves in finding a variational prior, which is difficult for graph data. To address this challenge, the authors propose to use MI neural estimation (MINE)~\cite{belghazi2018mine} to estimate the MI between the encoded subgraph and the input data. The results in~\cite{li2024task} demonstrate that GIB is more efficient than baseline methods on graph data inference tasks.



\subsubsection{Case Study}
\begin{figure}[t]
	\centering
        \includegraphics[width=0.9\linewidth]{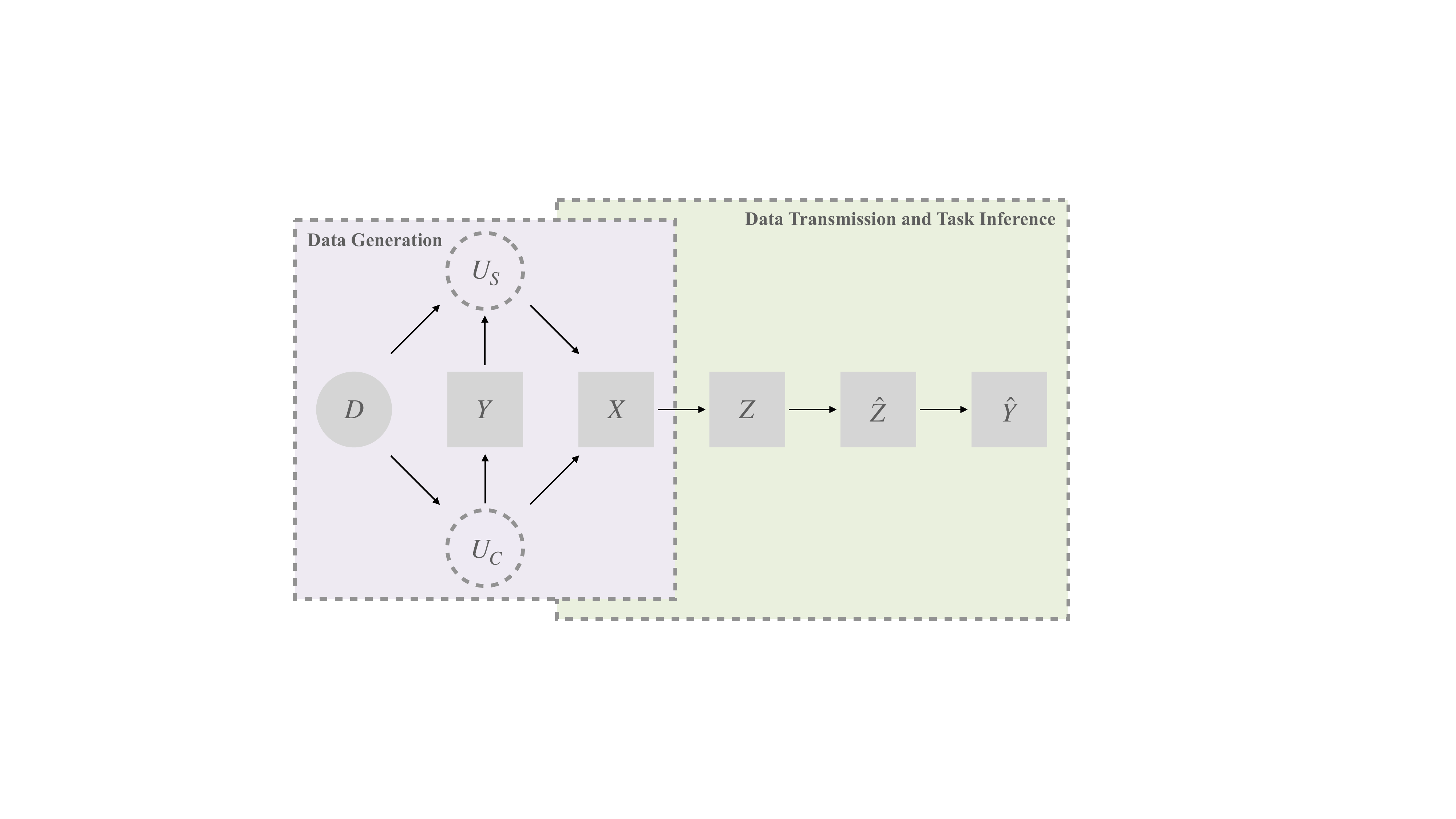}
        \caption{The PGM of the IFE-based task-oriented communication system.}
        \label{fig:pgm}
\end{figure}

\begin{figure}[t]
	\centering
        \includegraphics[width=0.9\linewidth]{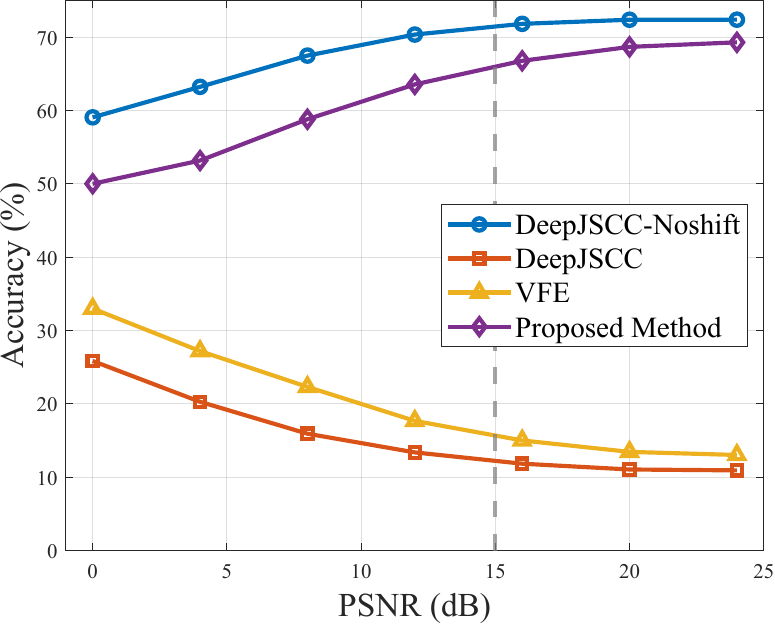}
        \caption{The causal feature inference accuracy on Colored-MNIST of different methods. All methods are trained on $\textrm{PSNR}_{\textrm{train}}=15\textrm{ dB}$.}
        \label{fig:acc}
\end{figure}

We take IFE as a case study to demonstrate how to use MI-based objective to design task-oriented communication systems. We begin by establishing the PGM among input data $X$, ground truth $Y$, causal features $U_C$, spurious features $U_S$, domain variable $D$, and transmitted/received signal $Z,\hat{Z}$, as shown in Fig.~\ref{fig:pgm}. The arrows in the PGM represent the dependencies and information flow among variables. Based on this PGM and $d$-separation principle, we can easily find that the MI between $Y$ and $D$ is zero when only conditioned on the causal feature $U_C$. This is because the domain variable $D$ is independent of the ground truth $Y$ when the causal feature $U_C$ is known. Therefore, in the objective function design, we can penalize the MI between $Y$ and $D$ conditioned on the received feature $\hat{Z}$ to force the encoded feature to only contain the information about $U_C$. To further enable tractable calculation of this conditional MI, we can use Taylor expansion and variational approximation to estimate the MI term in the objective function, as illustrated in~\cite{li2024tackling}. We compare the performance of IFE with VFE~\cite{vfe} and deep joint source-channel coding (DeepJSCC) under data distribution shift scenario. As shown in Fig.~\ref{fig:acc}, IFE significantly outperforms VFE and DeepJSCC, demonstrating its robustness to input data distribution shifts and its ability to generalize well to unseen conditions.

\subsection{Transmission Optimization}
The efficient and reliable transmission is a fundamental challenge in task-oriented communication. In this section, we will take digital modulation and precoding design as two examples to illustrate how MI can be leveraged to optimize transmission strategies in task-oriented communication systems.

\subsubsection{Robust Transmission for Digital Modulation}
In task-oriented communication, optimizing systems solely for task-relevant information extraction neglects two critical practical challenges: robustness to dynamic channel conditions and compatibility with standardized digital modulation schemes. MI also provides a unified framework to address these challenges while maintaining efficient information transmission. By constructing a Venn diagram among input data, transmitted symbols, received symbols, and task labels, as shown in Fig.~\ref{fig:component}, the authors in~\cite{xie2022robust_IB} propose a RIB framework to maximize the coded redundancy and minimize the task-relevant information.

Unlike the approaches that treat robustness as a separate concern, RIB integrates channel resilience directly into its MI-based optimization. Specifically, it maximizes the MI between received symbols and tasks while constraining the MI between transmitted and received symbols—a measure of transmission rate. This formulation ensures that modulated symbols retain structured redundancy: sufficient information to recover task-critical features under channel distortions, but without extraneous data that would violate bandwidth or compatibility requirements.

Compatibility with digital modulation arises naturally from the role of MI in RIB. Digital symbols, being discrete, allow MI between transmitted and received signals to be estimated directly, avoiding approximations required in analog systems. This makes MI optimization computationally tractable and aligns it with the constraints of standardized modulation schemes (e.g., QAM, PSK). By designing features where task-relevant information overlaps with channel-robust representations, RIB ensures seamless integration with existing infrastructure while adapting redundancy to channel conditions—a significant advance over static error-correction methods.

Experimental results in~\cite{xie2022robust_IB} validate this MI-based approach. Under varying signal-to-noise ratios, RIB-based systems achieve higher accuracy and lower symbol error rates than methods optimizing for task performance or channel resilience in isolation. This demonstrates the ability of MIto harmonize conflicting objectives: it discards semantically irrelevant data while preserving the redundancy necessary for both task execution and channel adaptation. These insights generalize beyond modulation. The ability of MI to quantify and optimize robustness and compatibility positions it as a cornerstone for task-oriented communication systems operating in real-world environments. By treating these challenges as inseparable facets of an MI optimization problem, designers can avoid ad-hoc solutions and build systems inherently adaptable to both task requirements and physical channel dynamics.
\subsubsection{Precoding Design for MIMO Systems}
In a task-oriented communication system over multiple-input multiple-output (MIMO) multiple access channel, the MIMO precoding design plays a critical role in optimizing the transmission strategy. However, the precoding design in a task-oriented communication system not only depends on the distribution of wireless channel and noise, but also relys on the task dataset and the learning model. Thus the high dimensional channel matrix in a MIMO setting makes the end-to-end learning of precoding design challenging and inefficient. 

MI can be naturally integrated into the precoding design of such problem, as it can directly model the mutual dependence between learning module and communication module. By analyzing the PGM among input data from multiple devices and channel states, the authors in~\cite{cai2025end} propose a conditional MI-based optimization framework for precoding design in MIMO systems. Specifically, the authors formulate the joint design of feature encoding, MIMO precoding and inference as a conditional MI maximization problem, i.e., maximizing the MI between the received signal and task conditioning on the channel state. Through this formulation, they connect the channel distribution, the task dataset, and the precoding design. To further reduce the training cost, the authors decouple the optimization of the precoding design from the task-learning module. They propose to align both the task-learning module and the precoding design with the original MI maximization through maximal coding rate reduction algorithm.

Simulation results in~\cite{cai2025end} demonstrate that the MI-based precoding design outperforms traditional non-MI-based methods in both the final task performance and the training efficiency, highlighting the effectiveness of MI-based transmission optimization in task-oriented communication systems.

\subsection{Efficient Training}
Previous work always overlook the training communication overhead in task-oriented communication systems, especially when the transmitter and receiver are trained in a decentralized manner. Such decentralized training requires frequent exchanges of intermediate features and gradients, which can strain communication resources in bandwidth-limited environments. MI provides a principled approach to improve training efficiency by task-agnostic feature extraction and feature alignment across transceivers.

\subsubsection{MI-based Hybrid Framework for Single Transceiver Training} For efficient single transceiver training, a MI-based hybrid training framework is proposed in~\cite{li2025icc}, which can reduce communication overhead while enhancing the representation learning capability of the transceiver models~\cite{li2025icc}. This framework consists of two sequential stages: \textit{self-supervised} local MI maximization and \textit{supervised} task-aware MI maximization. In the first stage, the transmitter independently learns feature representations by maximizing the MI between two different transformations of the same input, processed through a shared encoder. This approach encourages the model to capture intrinsic semantic representations invariant to geometric transformations, ensuring \textit{task-agnostic} feature extraction. The self-supervised learning paradigm minimizes data annotation requirements and reduces initial communication overhead.

The second stage involves \textit{supervised} task-aware MI maximization, where the transmitter and receiver undergo joint fine-tuning by exchanging intermediate features and gradients. The objective is to maximize the MI between the encoded features and the task, allowing the model to refine representations in a task-specific manner. By leveraging knowledge from the self-supervised phase, this stage significantly reduces gradient updates and data transmission, optimizing the trade-off between computation and communication costs.

This hybrid approach enhances both communication efficiency and system adaptability, supporting incremental updates as new data arrives. Particularly in dynamic environments, it ensures continuous fine-tuning without requiring frequent full model parameter transfers. The integration of MI maximization at both task-agnostic and task-specific levels ensures that learned representations remain informative, adaptive, and efficient for task-oriented communication.

\subsubsection{MI-based Feature Alignment for Retraining-Free Cross-Transceiver Communication}

Current task-oriented communication systems are typically optimized in a point-to-point manner, leading to inconsistent feature representations across independently trained transceiver pairs. Variations in model initialization and optimization paths result in non-uniform feature spaces, making direct communication infeasible without fine-tuning or retraining. This limitation hinders scalability and adaptability in multi-user networks.

A retraining-free MI-based feature alignment framework~\cite{xie2024toward} addresses this issue by maximizing MI between received features, task, and shared anchor data points across transceiver pairs. These anchor data samples provide a common reference for aligning independently trained transceivers. The approach builds on two key assumptions: \textit{angle preservation}, ensuring relative feature orientations remain consistent across encoders, and \textit{linear invariance}, suggesting feature representations across transceivers can be mapped using a linear transformation.

To achieve feature alignment, MI-based techniques are applied to both the transmitter and receiver:

\begin{itemize}
  \item \textit{Transmitter-Side Alignment}: Based on the \textit{angle preserving} assumption, the transmitter encodes and transmits relative feature angles of input samples with respect to anchor data, enabling the receiver to utilize angle information rather than absolute feature values for inference.
  \item \textit{Receiver-Side Alignment}: Using the \textit{linear invariance} assumption, the receiver estimates a transformation matrix between received features and anchor features, maximizing MI with both the anchor data and task. This transformation can be computed using learning-based methods or closed-form solutions such as Least Squares (LS) and Minimum Mean Square Error (MMSE).
\end{itemize}

By performing MI maximization, this approach eliminates the need for full model retraining when introducing new transceivers, enabling seamless interoperability across independently trained pairs. The framework enhances scalability and generalization, making it particularly suitable for dynamic, large-scale networks.

\subsubsection{Case Study}
\begin{figure}
  \centering
  \subfloat[Non-alignment]{
      \label{subfig: case-1-1}
    \centering
    \includegraphics[width=0.9\linewidth]{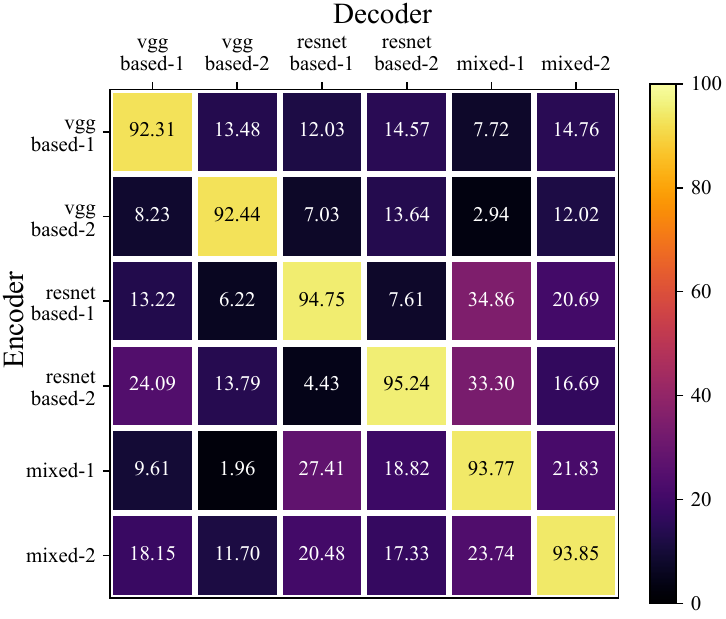}
  }
      
      \subfloat[Receiver-based alignment]{
      \label{subfig: case-1-2}
    \centering
    \includegraphics[width=0.9\linewidth]{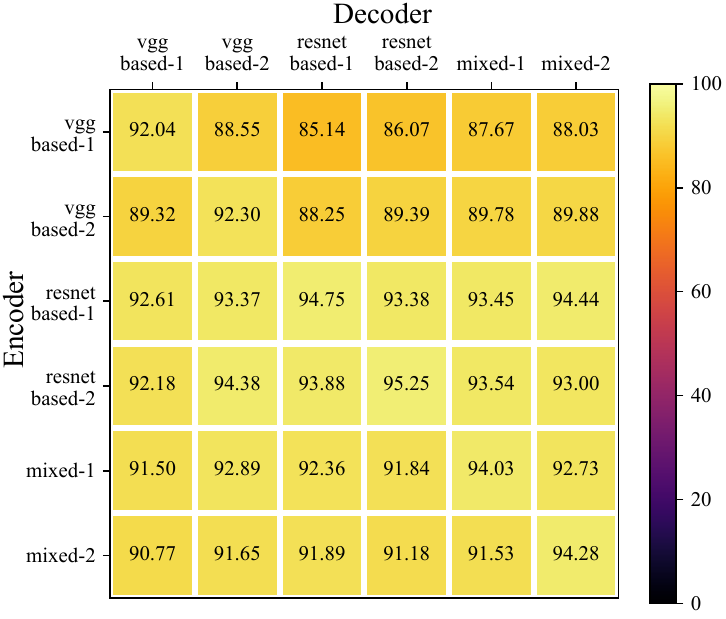}
  }
      \caption{The cross-transceiver task-oriented communication performance (a) without any feature alignment (Non-alignment) and (b) with receiver-side feature alignment (Receiver-side alignment) under the AWGN channels with $\textrm{SNR} =  10\textrm{ dB}$ on the SVHN dataset.
  }
      \label{fig: case-1}
\end{figure}

Figure~\ref{fig: case-1} demonstrates the effectiveness of receiver-side alignment of the MI-based cross-transceiver feature alignment in task-oriented communication. In the numerical simulation, three different neural network architectures are adopted as the backbones of encoder/decoder in transceivers and are trained independently within $i$-th transceiver, denoted as vgg-based-$i$, resnet-based-$i$, and mixed-$i$. 
The classification accuracy is used to evaluate the inference performance and a higher accuracy value indicates better performance.
In Figure~\ref{subfig: case-1-1}, it is observed that the diagonal entries exhibit satisfactory accuracy, while other entries fall below 20\%. It demonstrates the significant performance loss of direct cross-transceiver inference without any alignment. Figure~\ref{subfig: case-1-2} shows the performance of cross-transceiver task-oriented communication with the MI-based feature alignment method. It achieves the best performance and has
small performance gaps compared to the optimal performance. Note that the MI-based method not only works on the cross-transceiver inference between distinct training processes but also shows effectiveness for cross-architecture scenarios.

\subsection{Challenges and Limitations}
\label{subsec:challenges}
Although MI have shown promising results in task-oriented communication, there are still several challenges and limitations for this method. In this section, we will introduce several potential solutions to address these challenges when using MI as a tool in task-oriented communication.

\subsubsection{Intractable High-Dimensional Calculation}
Calculating MI in practical applications presents significant challenges due to its dependence on the closed-form joint distribution of two random variables. In most applications of MI, this joint distribution is either unknown or difficult to obtain. Even if the closed form is available, the high-dimensional integration required to evaluate MI remains computationally intractable. Consequently, MI-based objective optimization often relies on estimation or approximation techniques. A widely adopted approach is variational approximation, which introduces a simple, known variational distribution to approximate MI, as commonly seen in IB-related methods~\cite{vfe,shao2022task,li2024tackling,li2024task,xie2022robust_IB}. Beyond variational approximations, neural estimation methods have also been developed, including MINE~\cite{belghazi2018mine}, which provides a lower bound on MI, and Contrastive Log-ratio Upper Bound (CLUB)~\cite{cheng2020club}, which provides an upper bound on MI. These techniques are extensively utilized for system optimization in task-oriented communication. Further discussions on alternative bounds and theoretical insights into MI estimation can be found in~\cite{poole2019variational}.

\subsubsection{Invariance to Bijective Transformations}
MI inherently exhibits bijective invariance issue, meaning that MI remains unchanged under bijective transformations of the variables. Specifically, when MI is used to evaluate the relationship between feature representations and input/output variables, features that are transformed through bijective mappings (e.g., sigmoid or exponential functions) yield the same MI value despite potentially having significantly different performance in downstream learning tasks. This invariance issue can lead to suboptimal feature representations and deteriorate the learning performance of task-oriented communication systems. Potential solutions are to incorporate MI as a regularization term in the learning objective or use some information measures that defined on the feature space~\cite{xu2024neural}.

\section{Future directions and open issues}
\label{sec:future}
Previous sections have discussed the roles, applications and challenges of MI in task-oriented communication systems, several open issues and future directions are highlighted as follows.

\subsection{Multi-Task Learning in Task-Oriented Communication}
In real-world deployments, task-oriented communication systems often need to support multiple downstream tasks simultaneously (e.g., image recognition, speech translation, and control signal prediction in autonomous systems). A key challenge is ensuring that the transmitted features contain enough relevant information for all tasks without redundancy. To address this issue, further research is needed to develop multi-task learning strategies to remove task-specific redundancy to support multiple tasks while minimizing transmission overhead. A potential solution is task-conditional MI regularization, i.e., instead of treating all features equally, MI can be used to separate shared and task-specific representations, ensuring that common features are efficiently transmitted while task-unique information is selectively encoded.

\subsection{Multi-Modal Learning in Task-Oriented Communication}
Many task-oriented communication systems operate in multi-modal settings, where different types of sensory inputs (e.g., audio, video, text, radar, LiDAR) must be efficiently encoded and transmitted for complex tasks such as autonomous driving, remote medical diagnosis, or augmented reality. A key challenge is how to optimally fusing multiple modalities while avoiding redundant transmission. Future research should explore multi-modal MI optimization strategies to leverage the unique information of each modality and the shared information across modalities to enhance system performance.

\subsection{Power Control in Task-Oriented Communication}
In task-oriented communication, the importance of different features varies depending on the downstream task, making conventional power allocation schemes suboptimal. A key challenge is ensuring that the most task-relevant features are transmitted reliably without unnecessary energy expenditure. To address this problem, future research should explore MI-based power allocation, where MI is used as the measurement to model the mutual dependence between feature importance and transmission power. By allocating higher transmission power to semantically critical features and reducing power for redundant or less informative components, task-oriented power control can improve both energy efficiency and task performance under varying channel conditions.

\section{Conclusions}
The ability to unify communication and learning under a single, information-theoretic framework makes MI a powerful tool for designing task-oriented communication systems. Unlike heuristic or task-specific approaches, MI provides a principled way to quantify relevance, redundancy, and robustness across system components, enabling end-to-end optimization with both interpretability and generalizability. Building on this foundation, this paper provided a systematic overview of how MI-based approaches can be leveraged for feature encoding, transmission optimization, and efficient training in task-oriented communication. We explored key techniques and demonstrated their effectiveness in improving the reliability and efficiency of communication systems. Additionally, we elaborated on major challenges associated with MI-based methods and outlined potential solutions. Finally, we discussed open research directions, highlighting the promise of MI-driven optimization in advancing future intelligent communication systems.
\linespread{1.0}{
\bibliographystyle{./bibtex/IEEEtran}
\bibliography{./bibtex/IEEEabrv,ref}

\begin{thebibliography}{10}
\providecommand{\url}[1]{#1}
\csname url@samestyle\endcsname
\providecommand{\newblock}{\relax}
\providecommand{\bibinfo}[2]{#2}
\providecommand{\BIBentrySTDinterwordspacing}{\spaceskip=0pt\relax}
\providecommand{\BIBentryALTinterwordstretchfactor}{4}
\providecommand{\BIBentryALTinterwordspacing}{\spaceskip=\fontdimen2\font plus
\BIBentryALTinterwordstretchfactor\fontdimen3\font minus \fontdimen4\font\relax}
\providecommand{\BIBforeignlanguage}[2]{{%
\expandafter\ifx\csname l@#1\endcsname\relax
\typeout{** WARNING: IEEEtran.bst: No hyphenation pattern has been}%
\typeout{** loaded for the language `#1'. Using the pattern for}%
\typeout{** the default language instead.}%
\else
\language=\csname l@#1\endcsname
\fi
#2}}
\providecommand{\BIBdecl}{\relax}
\BIBdecl

\bibitem{shannon1948mathematical}
C.~E. Shannon, ``A mathematical theory of communication,'' \emph{Bell Syst. Tech. J.}, vol.~27, no.~3, pp. 379--423, 1948.

\bibitem{vfe}
J.~Shao, Y.~Mao, and J.~Zhang, ``Learning task-oriented communication for edge inference: An information bottleneck approach,'' \emph{IEEE J. Sel. Areas Commun.}, vol.~40, no.~1, pp. 197--211, Jan. 2022.

\bibitem{xie2021deep}
H.~Xie, Z.~Qin, G.~Y. Li, and B.-H. Juang, ``Deep learning enabled semantic communication systems,'' \emph{IEEE Trans. Signal Process.}, vol.~69, pp. 2663--2675, 2021.

\bibitem{gunduz2022beyond}
D.~G{\"u}nd{\"u}z, Z.~Qin, I.~E. Aguerri, H.~S. Dhillon, Z.~Yang, A.~Yener, K.~K. Wong, and C.-B. Chae, ``Beyond transmitting bits: Context, semantics, and task-oriented communications,'' \emph{IEEE J. Sel. Areas Commun.}, vol.~41, no.~1, pp. 5--41, 2022.

\bibitem{shao2022task}
J.~Shao, Y.~Mao, and J.~Zhang, ``Task-oriented communication for multidevice cooperative edge inference,'' \emph{IEEE Trans. Wireless Commun.}, vol.~22, no.~1, pp. 73--87, Jul. 2022.

\bibitem{xie2022robust_IB}
S.~Xie, S.~Ma, M.~Ding, Y.~Shi, M.~Tang, and Y.~Wu, ``Robust information bottleneck for task-oriented communication with digital modulation,'' \emph{IEEE J. Sel. Areas Commun.}, vol.~41, no.~8, pp. 2577--2591, Jun. 2023.

\bibitem{cai2025end}
C.~Cai, X.~Yuan, and Y.-J.~A. Zhang, ``End-to-end learning for task-oriented semantic communications over mimo channels: An information-theoretic framework,'' \emph{IEEE J. Sel. Areas Commun.}, 2025.

\bibitem{li2024tackling}
H.~Li, J.~Shao, H.~He, S.~Song, J.~Zhang, and K.~B. Letaief, ``Tackling distribution shifts in task-oriented communication with information bottleneck,'' \emph{arXiv preprint arXiv:2405.09514}, 2024.

\bibitem{li2024task}
S.~Li, Y.~Wang, S.~Guo, and C.~Feng, ``Task-oriented communication for graph data: A graph information bottleneck approach,'' \emph{IEEE Trans. Cogn. Commun. Netw.}, 2024, early access.

\bibitem{belghazi2018mine}
M.~I. Belghazi, A.~Baratin, S.~Rajeswar, S.~Ozair, Y.~Bengio, A.~Courville, and R.~D. Hjelm, ``Mine: mutual information neural estimation,'' \emph{arXiv preprint arXiv:1801.04062}, 2018.

\bibitem{li2025icc}
H.~Li, H.~Zhao, H.~He, S.~Song, J.~Zhang, and K.~B. Letaief, ``Remote training in task-oriented communication: Supervised or self-supervised with fine-tuning?'' in \emph{Proc. IEEE Int. Conf. Commun. (ICC)}, Montreal, Canada, Jun. 2025.

\bibitem{xie2024toward}
S.~Xie, H.~He, S.~Song, J.~Zhang, and K.~B. Letaief, ``Toward real-time edge {AI}: Model-agnostic task-oriented communication with visual feature alignment,'' \emph{arXiv preprint arXiv:2412.00862}, 2024.

\bibitem{cheng2020club}
P.~Cheng, W.~Hao, S.~Dai, J.~Liu, Z.~Gan, and L.~Carin, ``Club: A contrastive log-ratio upper bound of mutual information,'' in \emph{International conference on machine learning}.\hskip 1em plus 0.5em minus 0.4em\relax PMLR, 2020, pp. 1779--1788.

\bibitem{poole2019variational}
B.~Poole, S.~Ozair, A.~Van Den~Oord, A.~Alemi, and G.~Tucker, ``On variational bounds of mutual information,'' in \emph{Int. Conf. Mach. Learn.}\hskip 1em plus 0.5em minus 0.4em\relax PMLR, 2019, pp. 5171--5180.

\bibitem{xu2024neural}
X.~Xu and L.~Zheng, ``Neural feature learning in function space,'' \emph{J. Mach. Learn. Res.}, vol.~25, no. 142, pp. 1--76, 2024.

\end{thebibliography}
}
\end{document}